\documentclass[runningheads]{llncs}
\usepackage[T1]{fontenc}
\usepackage{graphicx,verbatim}
\usepackage{hyperref}
\usepackage{color}

\usepackage{ifthen}
\usepackage{amsmath, amsfonts, amssymb}
\usepackage{float}
\usepackage{booktabs}
\usepackage{subfig}
\usepackage{enumitem}
\usepackage{bbm}
\usepackage{gensymb}
\usepackage{mathtools}
\usepackage{multirow}
\usepackage{multicol}
\usepackage{xspace}
\usepackage[square,numbers,sort&compress,sectionbib]{natbib}

\makeatletter
\DeclareRobustCommand\onedot{\futurelet\@let@token\@onedot}
\def\@onedot{\ifx\@let@token.\else.\null\fi\xspace}

\def\etal{\emph{et al}\onedot}

\newcommand{\realxray}{\textsc{RealXray}\xspace}
\newcommand{\realctdrr}{\textsc{RealCT-DRR}\xspace}
\newcommand{\synthxray}{\textsc{DiffXray}\xspace}
\newcommand{\synthxraycloud}{\textsc{DiffXray-Cloud}\xspace}

\newcommand{\synthctdrr}{\textsc{DiffCT-DRR}\xspace}

\newcommand{\realctdrrcloud}{\textsc{RealCT-DRR-Cloud}\xspace}

\begin{document}
%
\title{2D Versus 3D Diffusion for \emph{In Silico} Training of Interventional X-ray AI Models}
\titlerunning{2D Versus 3D Diffusion for \emph{In Silico}}

\author{Sampath Rapuri\inst{\ast 1}\and
Jeremy Ko\inst{\ast 1}\and
Benjamin~D.~Killeen\inst{\ast 1}\and
Russell H.~Taylor\and
Mathias Unberath\inst{1}
}
\authorrunning{Rapuri et al.}
%
\institute{*Equal contribution.\\
Johns Hopkins University, Baltimore, MD 21218, USA\\
\email{\{killeen, srapuri1, jko26, rht, unberath\}@jhu.edu}}

\maketitle              
\begin{abstract}
  The ability to synthesize realistic X-ray images has catalyzed the development of AI models for X-ray image-guided procedures, which otherwise suffer from a lack of available annotated data.
  Prior work has demonstrated the effectiveness of mechanistic simulation of digitally reconstructed radiographs (DRRs) as a training data source for a myriad of tasks, including segmentation and anatomical landmark detection, with comparable or superior performance to real data training.
  However, mechanistic DRR synthesis still relies on the availability of annotated high-resolution anatomical models. Deriving these from CT images of real patients or specimens imposes an undesirable bottleneck on data quantity and variability. In this work, we explore two methods for synthesizing training data: (1) a 3D conditional latent diffusion model that generates CT volumes to use as inputs for mechanistic DRR generation without real, 3D anatomical models, and (2) a view-conditioned 2D diffusion model that produces synthetic X-rays. In controlled experiments, we demonstrate that synthetic 2D diffusion-based X-rays can be used to train an anatomical landmark detection model that generalized to real X-ray images with performance rivaling that of a model trained on real X-ray images. Thus, we provide preliminary evidence that  synthetic, 2D diffusion-based training data can substitute for real X-ray data, identifying a promising avenue towards generating large, diverse datasets for training robust AI models in interventional X-ray imaging.

  
\end{abstract}

\section{Introduction}

\begin{figure}
  \centering
  \includegraphics[width=\textwidth]{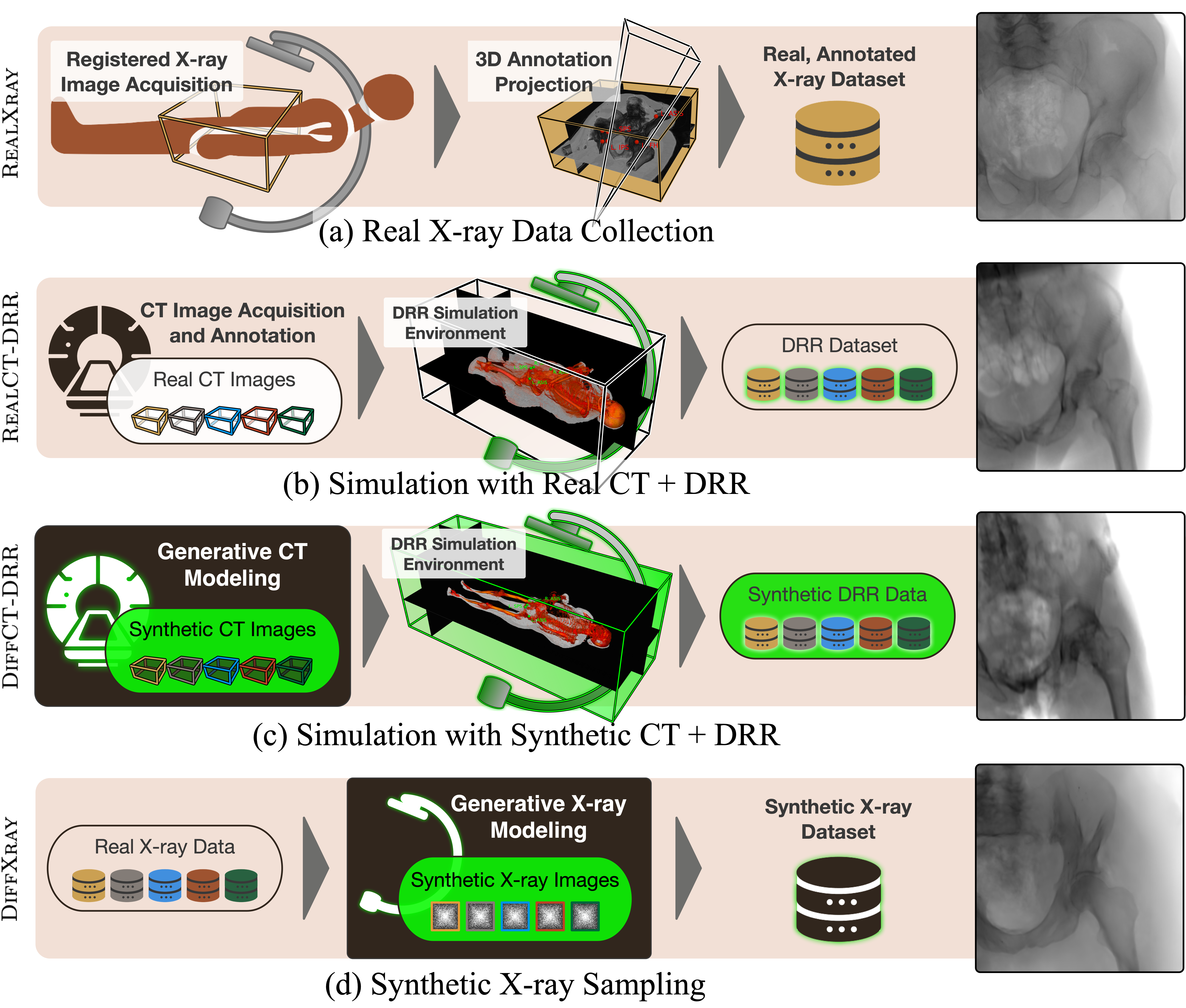}
  \caption{
    \textbf{Paradigms for obtaining X-ray training data.} (a) Real data collection (\realxray) requires 2D/3D registration with a CT image from which annotations can be projected~\cite{grupp2020automatic}.\ (b) DRR-based simulation (\realctdrr) facilitates large-scale, highly controllable data generation from real CT volumes~\cite{gao2023synthetic}.\ (c) The proposed 3D diffusion framework synthesizes CT images with a latent DM, from which DRRs can be projected (\synthctdrr). (d) The proposed 2D diffusion framework (\synthxray) generates realistic X-ray images aligned with segmentations and landmarks.
  }
  \label{fig:overview}
\end{figure}

\emph{In silico} simulation is a powerful tool for the development of artificial intelligence (AI) models in medical imaging~\cite{killeen2023silico}. This is particularly true in the context of interventional X-ray imaging, where neither images nor the desired annotations are generated during routine care~\cite{unberath2018deepdrr}. The ability to synthesize realistic X-ray images alongside ground truth annotations has catalyzed the development of AI models for image-guided procedures, with downstream applications including segmentation~\cite{zhang2020drr4covid,killeen2025fluorosam}, anatomical landmark detection,\cite{gao2023synthetic,bier2018xray} automated C-arm positioning~\cite{kausch2020toward, kausch2021carm}, surgical phase recognition~\cite{killeen2023pelphix}, intra-operative planning~\cite{killeen2023autonomous}, and intelligent interfaces~\cite{killeen2024take,killeen2025intelligent}. The predominant approach in this area, as proposed by Gao \etal~\cite{gao2023synthetic}, is to use real CT images as the patient model for synthesizing digitally reconstructed radiographs (DRRs), with ground truth segmentations, landmarks, or other annotation data projected from the 3D volume to the 2D DRR. In this approach, which we refer to as \realctdrr, domain randomization techniques enable training of AI models on synthetic data that overcome the sim-to-real gap, achieving comparable or superior performance to those trained on real images or models trained on images processed using supervised domain adaptation, e.\,g., via generative adversarial networks (GANs)~\cite{gao2023synthetic}. However, the success of this approach is predicated on the availability of annotated CT images from real patients or specimens, imposing an undesirable bottleneck on data quantity and variability~\cite{ricci2022addressing}.

Recently, diffusion models (DMs)~\cite{ho2020denoising} have emerged as a powerful tool for generating training data~\cite{khader2023denoising, guo2024maisi}. Their success in controllable image synthesis~\cite{guo2024maisi,weber2023cascaded,huangchest,killeen2025towards} raises the question of whether they can be used for \emph{in silico} training of AI models, where the data generation process outputs not only realistic images but also -- through conditioning -- the corresponding annotations, in a more flexible, scalable manner than \realctdrr. In this work, we examine the application of DMs to X-ray image and annotation synthesis under two novel, competing paradigms, focusing on pelvic X-ray anatomical landmark detection as a motivating task, as in~\cite{gao2023synthetic}. The first paradigm, which we refer to as \synthctdrr, uses a 3D DM~\cite{killeen2025towards} to generate synthetic CT images, which are automatically segmented and annotated with landmarks following~\cite{killeen2023pelphix}. The second novel paradigm, which we refer to as \synthxray, uses a 2D DM to generate X-ray images conditioned on plausible organ segmentations, such as can be obtained from a computational human phantom~\cite{segars2013population}. Fig.~\ref{fig:overview} illustrates the differences between these two paradigms and the prior approaches of \realxray and \realctdrr. 
In line with previous work~\cite{gao2023synthetic}, we find that \realctdrr performs exceptionally strongly given a sufficiently large and diverse set of poses with an average $\ell_2$ error of $4.37$ mm on a withheld test subject's real X-ray images. However, \synthxray enables comparable performance for synthetic X-ray generation with an average $\ell_2$ error of $4.84$ mm on the same withheld X-ray images, and \synthctdrr performs comparably to \realctdrr in a precisely matched experimental setup. In demonstrating the effectiveness of \synthxray in landmark detection, we validate an entirely \emph{in silico} paradigm of training AI models for interventional X-ray imaging. 







\section{Methods}

The following sections describe the two novel paradigms, \synthctdrr{} and \synthxray, as well as the landmark detection model trained on these data. We also describe the datasets used in our experiments and the evaluation metrics.

\subsection{\synthctdrr{}: DRRs from 3D Diffusion Modeled CT Images}

The \synthctdrr{} framework uses a 3D latent diffusion model to generate synthetic CT images, which are then used to synthesize DRRs. Our goal is to facilitate training for an automated landmark detection algorithm, which requires corresponding images, organ segmentations, and 2D landmark annotations. In our experiments, we use the full-body CT diffusion model from Killeen \etal~\cite{killeen2025towards}, a 3D DM that shows good realism in terms of both low-level visual fidelity and global anatomical realism. Crucially, this realism is sufficient to be compatible with robust multi-organ segmentation tools~\cite{wasserthal2023totalsegmentator}, from which we obtain surface meshes for the pelvis with marching cubes. Following \cite{killeen2023pelphix}, we automatically propagate 3D landmark annotations from a statistical shape model (SSM) to the synthetic CT, which are projected for each image to obtain the 2D annotations.

In our experiments, we use the registered CT and X-ray images from Grupp \etal~\cite{grupp2020automatic} as the real data source, which contains 366 X-ray images from six cadaveric specimens, each with a corresponding CT volume. To enable a fair comparison with other data paradigms, we generate DRRs from six distinct synthetic CT images that are precisely matched to the real X-ray poses. For each real specimen $r$, we compute a rigid transformation $\mathbf{F}_{r}^{s}$ that minimizes the mean distance between the landmarks in $r$ and the landmarks in the synthetic CT $s$. We then choose a synthetic CT from among 100 samples with the lowest mean distance to each real specimen. If $\mathbf{P}_r$ is the camera projection matrix for a real X-ray, we can generate a corresponding DRR with $\mathbf{P}_s = \mathbf{P}_r \cdot \mathbf{F}_{r}^{s}$, as shown in Fig.~\ref{fig:synthctdrr-frames}. The resulting DRRs are then used to train a landmark detection model, which we evaluate on real X-ray images from a withheld dataset.

\subsection{\synthxray{}: 2D X-ray Image Diffusion}

\newcommand{\figheight}{0.47\textwidth}
\begin{figure}[t]
  \centering
  \subfloat[\synthctdrr Frames]{
    \includegraphics[height=\figheight]{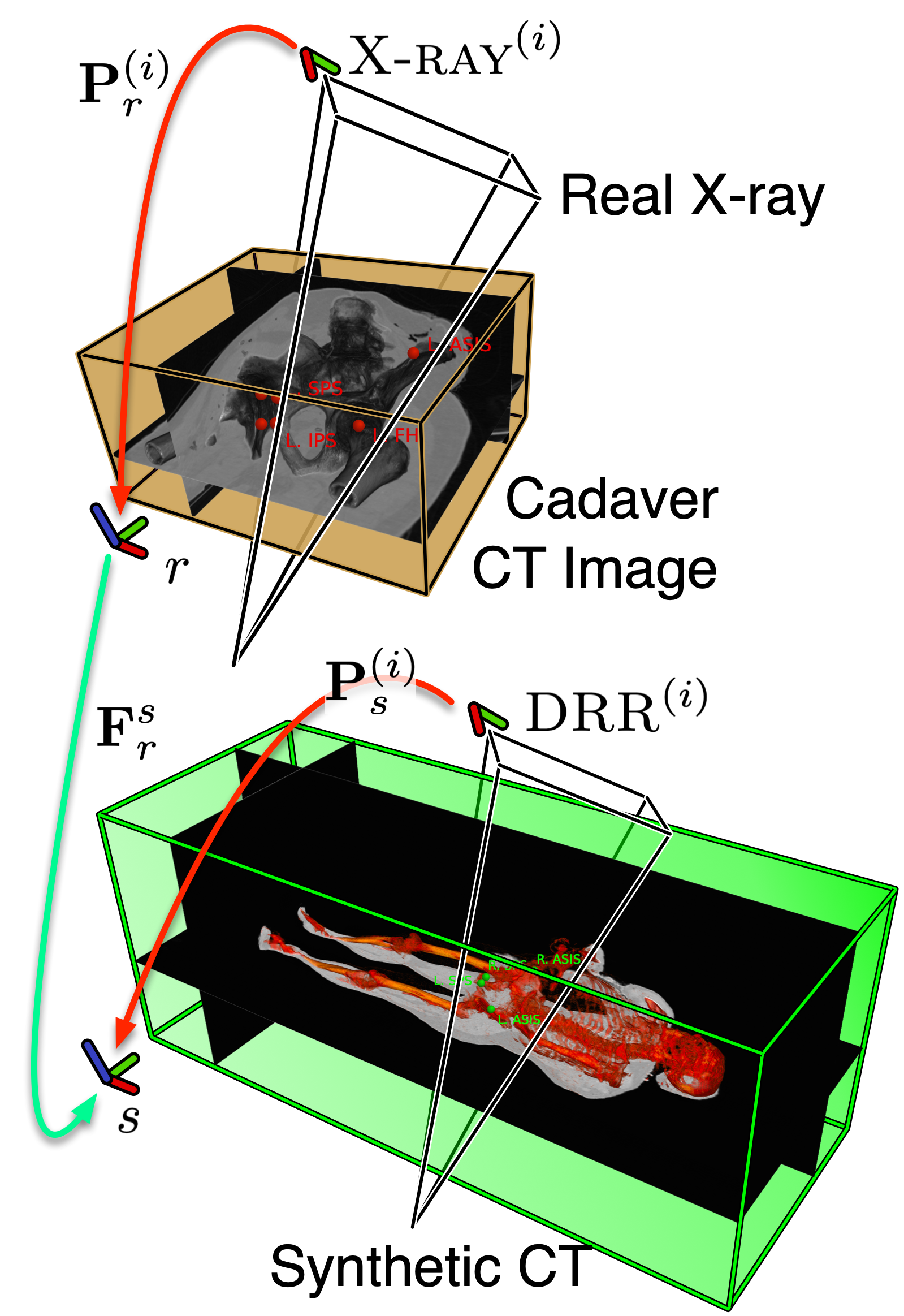}
    \label{fig:synthctdrr-frames}
  }
  \hfill
  \subfloat[\synthxray Training and Inference]{
    \includegraphics[height=\figheight]{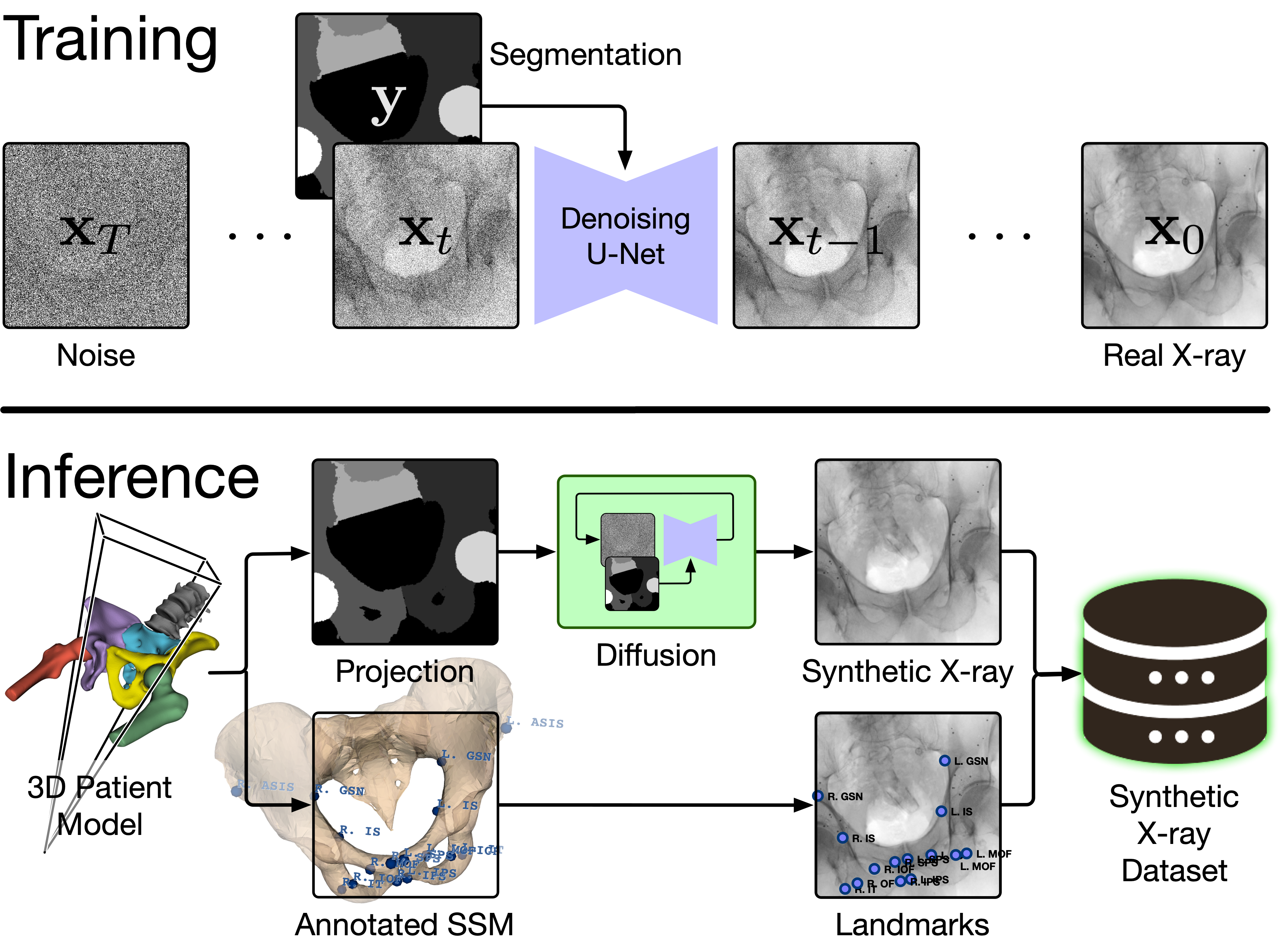}
    \label{fig:synthxray-training}
  }
  \caption{\protect\subref{fig:synthctdrr-frames} In our experiments, we generate precisely matched datasets with the DeepFluoro dataset~\cite{grupp2020automatic} by rendering DRRs from synthetic CT images which have been registered to the real cadaver CT images and each X-ray image. \protect\subref{fig:synthxray-training} The \synthxray{} framework, on the other hand, conditions a 2D diffusion model on segmentations which have been projected from a 3D patient model, with landmarks from the statistical shape model (SSM) in~\cite{killeen2023pelphix}.
  }
  \label{fig:data-generation}
\end{figure}

The goal of the \synthxray framework is to generate a dataset of synthetic X-ray images and landmarks with precisely controlled poses in one-to-one correspondence with our real dataset. As shown in Fig.~\ref{fig:synthxray-training}, in the \synthxray framework, a 2D DM is trained on the dataset of real X-ray images and corresponding segmentations \cite{killeen2023pelphix, konz2024anatomicallycontrollablemedicalimagegeneration}.  During training, we apply rotations of up to 20 degrees and random crops by up to 12\% of the image size, and mean squared error loss is used throughout. The DM was trained for 400 epochs with a batch size of 1 using a cosine learning rate scheduler with a learning rate of $\eta_0 = 2\times 10^{-5}$ on an NVIDIA Titan RTX GPU. In order to generate the synthetic X-ray dataset, segmentations are projected through a 3D patient model, and the projected segmentations condition the generation of synthetic X-rays by the DM. Landmark locations are obtained by projecting the landmarks from an annotated SSM registered to the patient model. In our experiments, we use meshes obtained from segmentations of cadaver CTs, but a computational human phantom such as in Segars~\etal~\cite{segars2013population} would also be sufficient and avoid the need for real CT images. By projecting through the same poses as the dataset of real X-ray images, we generate a synthetic X-ray dataset annotated with landmarks in one-to-one pose correspondence to our dataset of real images.


\subsection{Landmark Detection Model}


Following~\cite{killeen2023autonomous,gao2023synthetic,killeen2023pelphix}, our anatomical landmark detection model uses a Swin U-Net architecture~\cite{cao2021swin} with combined segmentation and heatmap output, with an input image size of $224\times 224$. DICE loss and NCC loss are weighted equally for the segmentation and heatmap outputs, respectively. For the cross-validation experiments, we train the model to convergence (5k epochs) with a batch size of 4 using the AdamW optimizer with a cosine-annealing learning rate scheduler with warm restarts, with $\eta_0 = 2\times 10^{-4}$, using a NVIDIA RTX 3090 GPU. To enable sim-to-real transfer learning, we use the domain randomization pipeline from~\cite{killeen2025fluorosam}, following Gao \etal~\cite{gao2023synthetic}, consistently across all experiments.




\section{Experiments} 

We evaluate each paradigm for training anatomical landmark detection models following the methodology in Gao \etal~\cite{gao2023synthetic}. The DeepFluoro dataset~\cite{grupp2020automatic} consists of 366 fluoroscopic images from 6 cadaveric specimens, registered to CT volumes with segmentations of the semi-pelvises, sacrum, femurs, and vertebrae. We conduct two types of experiments. In \emph{precisely matched cross-validation} experiments, DRRs and synthetic X-rays are generated to precisely match the views included in the real DeepFluoro images. Because of DeepFluoro's limited size, 6-way cross-validation is conducted. In every experiment, the model was evaluated on real X-ray images from a test set withheld during training of the landmark detection model. For \synthxray, the 366 images are likewise generated in a 6-fold manner, recreating the projections for each specimen using a DM trained on the other 5.

In the (2) ``cloud'' experiment, we compare the performance of \synthxray{} to the precisely matched experiments, in the scenario where extra training data can be stochastically sampled from arbitrary views, since this scalability is the primary benefit of generative modeling. Given each patient's existing set of views in spherical coordinates, we sample 500 views per patient by drawing uniformly from $\pm$2.5$\degree$ perturbations around each original viewpoint.

\begin{figure}[htbp]
  \centering
  \includegraphics[width=\linewidth]{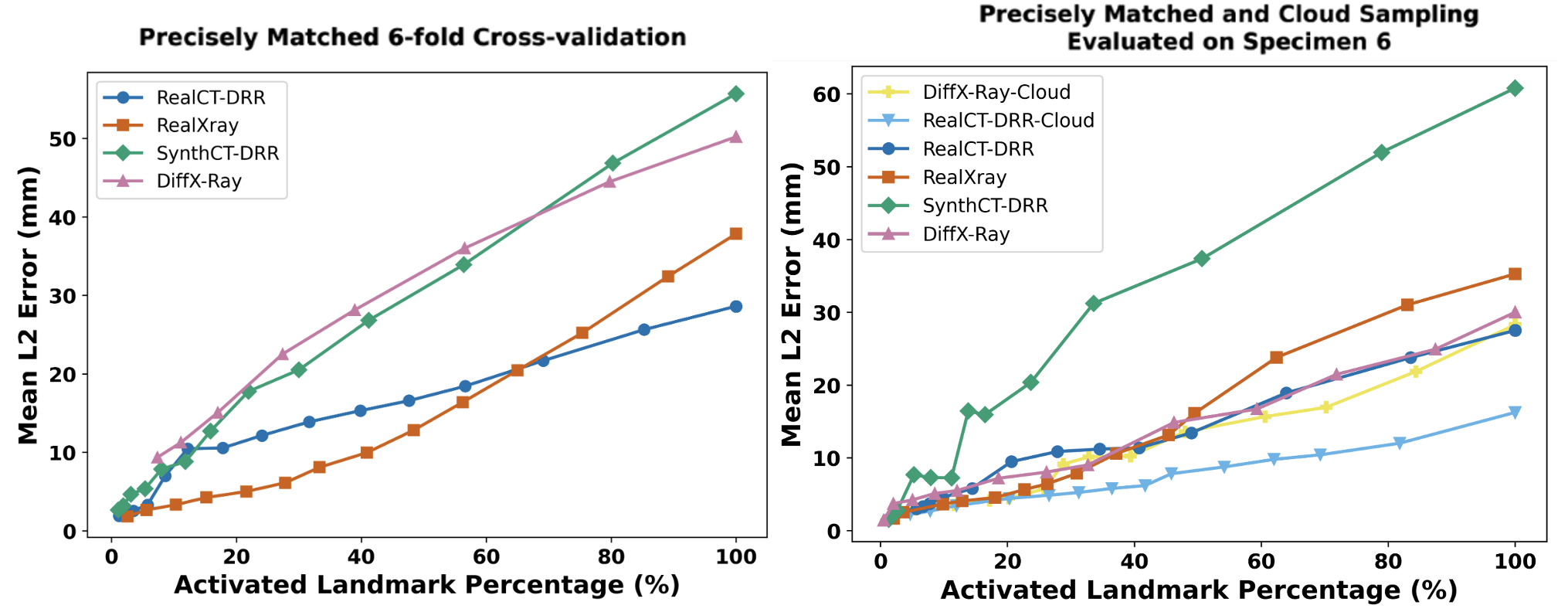}
  \caption{(a) Performance comparison of the four generative methods through the precisely matched cross-validation experiment. In these data-poor environments, the \synthxray framework performs poorly, but in (b) when used to increase the available training data through cloud sampling (c), it meaningfully enhances performance for the most challenging specimen, outperforming all other paradigms.}
  \label{fig:results}
\end{figure}

We present the anatomical landmark detection performance of our models for each data paradigm in Fig.~\ref{fig:results}. Like Gao \etal~\cite{gao2023synthetic}, we determine which landmarks to include based on the normalized cross-correlations (NCCs) between the predicted heatmaps ($\hat{h}$) and the ground truth ($h$) and threshold these landmarks at $\mathrm{NCC}(\hat{h}, h) > \phi$, decreasing $\phi$ to activate additional landmarks and plot the mean $\ell_2$ error for each threshold, averaged across each fold. A perfect model would demonstrate a 0 mm error at 100\% activation. As can be seen in Fig.~\ref{fig:results}a, there is no clear advantage between \realctdrr and \realxray, but \synthctdrr and \synthxray perform poorly. At first glance, this is at odds with results in Table~\ref{tab:landmark-errors}, where \synthxray achieves the lowest $\ell_2$ error of $4.40$ $\pm$ $1.87$ mm at a high NCC threshold. This reflects the precision of the model predictions at a high confidence and its rapid degradation with less confident predictions.



\vspace{-1em}  

In a data-rich environment, though, the advantage of \synthxray is apparent. Fig.~\ref{fig:results}b includes the $\ell_2$ error on DeepFluoro's specimen 6, a particularly challenging case due to its C-arm angle distribution. As can be seen, with additional variation from small perturbations in the C-arm angle, \realctdrrcloud results in the highest performance at $0.40$ and $0.20$ NCC (see Table~\ref{tab:landmark-errors}). However, a model trained on \synthxraycloud actually improves upon \realctdrrcloud at a high NCC threshold with an average $\ell_2$ error of $4.47\pm 1.49$ mm.

\begin{table}[htbp]
  \centering
  \caption{Landmark detection $\ell_2$ error, with $\mu \pm \sigma$ (95\% CI) across thresholds.}
  \label{tab:landmark-errors}
  \begin{tabular}{llrrr}
  \toprule
  ~~~ && \multicolumn{3}{c}{$\ell_2$ Error} \\
  \cmidrule(lr){3-5}
  && \multicolumn{1}{c}{0.60 NCC} & \multicolumn{1}{c}{0.40 NCC} & \multicolumn{1}{c}{0.20 NCC} \\
  \midrule
\multicolumn{5}{l}{\textit{Precisely matched (6‐fold Cross‐validation)}} \\
  & RealCT‐DRR      & $7.83 \pm 25.67$ (2.91)  & $12.32 \pm 27.33$ (1.91) & $17.92 \pm 30.53$ (1.54) \\
  & RealXray        & $4.89 \pm 6.42$  (0.51)  & $ 9.50 \pm 19.61$ (1.10) & $18.94 \pm 35.57$ (1.57) \\
  & SynthCT‐DRR     & $5.82 \pm 13.72$ (2.17)  & $12.36 \pm 25.82$ (2.49) & $23.64 \pm 38.73$ (2.33) \\
  & DiffX‐Ray       & $4.40 \pm 1.87$  (0.52)  & $12.17 \pm 26.88$ (3.63) & $25.40 \pm 37.93$ (2.68) \\
\multicolumn{5}{l}{\textit{Precisely matched (Specimen 6)}} \\
  & RealCT‐DRR      & $4.37 \pm 2.08$  (0.96)  & $11.08 \pm 18.72$ (5.35) & $12.70 \pm 15.08$ (3.17) \\
  & RealXray        & $4.38 \pm 1.84$  (0.63)  & $ 7.43 \pm  8.01$ (2.10) & $14.65 \pm 22.56$ (4.58) \\
  & SynthCT‐DRR     & $8.00 \pm 10.80$ (8.00)  & $16.59 \pm 33.32$ (14.60)& $23.80 \pm 41.08$ (12.28) \\
  & DiffX‐Ray       & $4.84 \pm 1.45$  (0.71)  & $ 7.59 \pm  4.11$ (1.17) & $16.80 \pm 19.72$ (3.72) \\
\multicolumn{5}{l}{\textit{Cloud experiments}} \\
  & RealCT‐DRR‐Cloud & $4.87 \pm 1.75$  (0.49)  & $ 6.02 \pm  2.40$ (0.54) & $ 9.73 \pm 13.16$ (2.42) \\
  & DiffX‐Ray‐Cloud  & $4.47 \pm 1.49$  (0.47)  & $10.12 \pm 22.11$ (5.42) & $15.09 \pm 26.55$ (4.98) \\
  \bottomrule
  \end{tabular}
\end{table}


\section{Discussion}

Because of data scarcity in medical imaging, generative AI models have proven valuable for augmenting or replacing real training data with synthetic samples, improving downstream performance~\cite{khader2023denoising}. For interventional X-ray imaging, the dominant paradigm so far has involved the mechanistic simulation of DRRs in large quantities for widely varied viewpoints. This is to contend with the high variability of images in the interventional setting, where changes to acquisition protocols, such as the angulation of a C-arm, result in distribution shifts that can severely impact performance. Nevertheless, there is significant potential for generative AI models to improve downstream performance for interventional X-ray image analysis, due to their high degree of realism. At the same time, conditional diffusion models offer a path forward due to their realism and high degree of controllability in the sense that by controlling the poses used to generate the conditioning segmentations, the poses of the diffused images can be completely controlled~\cite{konz2024anatomicallycontrollablemedicalimagegeneration, killeen2025towards}. Our goal in this paper has been to evaluate two competing paradigms for conditional diffusion-based data generation, with a focus on anatomical landmark detection in pelvic surgery.

Following prior work, our experiments evaluate each paradigm in a fair comparison. By precisely matching the viewpoints of generated images with real X-rays, one can ensure that the only difference in performance of the resulting landmark detection model comes from the difference in data domains between each framework. In this precisely matched setup, we find that \synthctdrr rivals the performance of \realctdrr at $0.60$ and $0.40$ NCC thresholds, indicating that \synthctdrr may hold potential as a completely synthetic alternative to \realctdrr. However, as seen in Table~\ref{tab:landmark-errors}, where \synthctdrr achieves comparable performance to \realctdrr at these NCC thresholds, there is a rapid decline in performance with less confident predictions at lower thresholds, highlighting a limitation of this data generation paradigm.

\synthxray rivals \realxray at high confidence predictions. Restricting the output of the \synthxray framework in this way, though, obscures its advantages: scalability and controllability. By conditioning images in the \synthxray framework from a wider variety of views -- even wider than in the training set for the DM -- it is possible to improve the performance of downstream landmark detection significantly, from $12.17\pm 26.88$ mm to $10.12\pm 22.11$ at a $0.40$ NCC threshold and $25.40\pm 37.93$ mm to $15.09\pm 26.55$ at a $0.20$ NCC threshold.

There are notable limitations to the results presented here. For one, there are few models available for CT image synthesis \cite{guo2024maisi}, conditional or otherwise, and the model from Killeen~\etal~\cite{killeen2025towards} has limited spatial resolution and conditioning signal. Mechanistic simulation is only as realistic as the underlying model. Although the current sim-to-real gap for \realctdrr can be overcome through domain randomization in many cases, this work shows that there is potential for meaningful data generation using 2D diffusion models.

\section{Conclusion}

We have shown that realistic simulation of X-ray images through the generation of DRRs from real CTs and diffused images is a viable alternative to real data for training downstream models for tasks like anatomical landmark detection. We quantified the performance of landmark detection models trained on different data generation paradigms through controlled experiments in which the distribution of pelvis poses remained constant across different synthetic data generation paradigms. We discovered promising sim-to-real transfer of models trained on \synthxray that rivals models trained on real data. We found that \synthctdrr outperforms \realctdrr across confident landmark predictions and rivals \realctdrr in performance at high confidence landmark predictions. Because the diffusion pipelines presented here do not necessitate any real data during inference when used to train downstream machine learning models, these two synthetic data pipelines hold much potential for training tasks that suffer from a lack of real data like interventional X-ray imaging. 

\bibliographystyle{splncs04}
\bibliography{references-short}


\end{document}